\documentclass[12pt]{article}
\usepackage{graphicx}
\usepackage[small]{caption}

\title{\textbf{A Cyclic Cosmological Model in the Landscape Scenario}}
\author{\textbf{Cheng-Yi Sun\footnote{cysun@mails.gucas.ac.cn}
and De-Hai Zhang\footnote{dhzhang@gucas.ac.cn}}\\
Department of Physics,\\
The Graduate School of The Chinese Academy of Sciences,\\
Beijing 10049, P.R.China.}

\begin{document}
\maketitle
\begin{abstract}
In \cite{KKLT}, KKLT give a mechanism to generate de Sitter vacua in
string theory. And the scenario, \emph{Landscape}, is suggested to
explain the problem of the cosmological constant. In this paper,
adopting a simple potential describing the \emph{landscape}, we
investigate the decay of the vacuum and the evolution of the
universe after the decay. We find that the big crunch of the
universe is inevitable. But, according to the modified Friedmann
equation in \cite{t0506218}, the singularity of the big crunch is
avoided. Furthermore, we find that this gives a cyclic cosmological
model.
\end{abstract}



\section{Introduction}
The data from the observation of the first year Wilkinson
Microwave Antisotropy Probe (WMAP) \cite{a0302207,a0302209} and
the observation of the SNe Ia \cite{Ia SuperNova} make us almost
be sure that the expansion of the universe is accelerating. The
simplest explanation is that a small but non-zero cosmological
constant, the de Sitter vacuum, dominates the present universe.
Recent years, great efforts basing on the string theory have been
paid to solve the cosmological constant problem and to construct a
complete process of the cosmological evolution. In\cite{KKLT}, a
mechanism, KKLT mechanism, is given to get de Sitter vacua in
string theory. And a scenario named \emph{Landscape} has been
suggested \cite{t0302219,t0408133,t0501082}. In this scenario it
is argued that string theory has a landscape of vacua. The
supersymmetric (SUSY) sector of the landscape has the zero vacuum
energy. The non-SUSY sector has a stochastic distribution of vacua
energies around the zero vacuum energy, where some vacua are de
Sitter vacua with positive vacuum energy and others are anti-de
Sitter vacua with negative vacuum energy. One of the de Sitter
vacua describes the present acceleration of our unverse. So the
cosmological constant, as a metastable de Sitter vacuum, must
decay into an anti-de Sitter vacuum. Unfortunately, the detailed
information of the landscape is absent. Here we simply take the
model in \cite{KL} to describe it.

On the other hand, the cyclic cosmological model has been suggested
as a radical alternative to inflation scenario
\cite{t0103239,t0111030}(For a short review, see
Ref.\cite{a0404480}). In this scenario the universe undergoes an
endless sequence of cosmic epochs each beginning with a `bang' and
ending in a `crunch'. It gives a whole process of the evolution of
the universe. In the current work, we find that the landscape
indicates a cyclic model naturally.

In this paper, we first assume that the de Sitter vacuum of the
potential suggested in \cite{KL} describes the present acceleration
of the universe, and then discuss the evolution of the universe
after the decay of the vacuum. We find that the contraction is
obtained inside the bubble which is materialized as the decay. So
the singularity of the big crunch is encountered. We know the
singularity is a long-standing issue in theoretical physics. Here,
we find, using the modified Friedmann equation in \cite{t0506218},
the singularity of the contraction is avoided and the bounce appears
as the end of the contraction. Finally we show this scenario gives a
cyclic cosmological model.

\section{de Sitter Vacua and anti-de Sitter Vacua in String Theory}

\begin{figure}
\centering
\renewcommand{\figurename}{Fig.}
\includegraphics{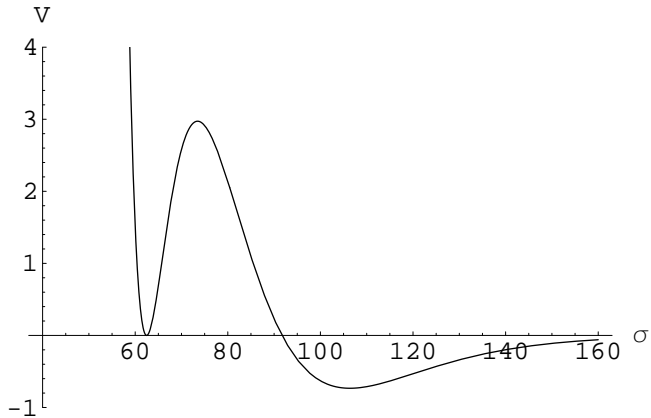}
\caption{The potential (\ref{dspotential}), multiplied by $10^{14}$,
for the values of the parameters $A=1$, $B=-1.03$, $a=2\pi/100$,
$b=2\pi/99$, and $W_0=-2\times10^{-4}$.\label{KLp}}
\end{figure}

In the theory of the $N=1$ supergravity, the potential is
\begin{equation}
\label{potential}
V=e^{K}\left(\sum_{a,b}G^{a\bar{b}}D_{a}W\overline{D_{b}W}-3|W|^{2}\right),
\end{equation}
where $a,b$ runs over all the modulus fields and $K$ is the
K\"{a}hler potential, $K=-3\ln [(\rho +\bar{\rho})]$. The volume
modulus $\rho$ is simply taken to be the real field
$\rho=\bar{\rho}=\sigma$. In the simplest KKLT model \cite{KKLT},
the superpotential $W$ is given by $W=W_{0}+Ae^{-a\rho }$. When
the potential is supplemented by a D-type contribution
$\frac{D}{\sigma ^{3}}$ from anti-D3 brane \cite{KKLT} or D7
branes \cite{t0309187}, a de Sitter minimum is found. In
\cite{KL}, this model is slightly modified by taking
\begin{equation}
W=W_{0}+Ae^{-a\rho }+Be^{-b\rho }, \label{superpotential}
\end{equation}
in order to be compatible with supersymmetry breaking and
inflation. Here, $W_{0}$ is a tree level contribution which arises
from the fluxes, $A,a$ and $b$ are positive constants, and $B$ is
a negative constant. Now the potential in equation
(\ref{potential}) is written as
\begin{eqnarray}
V&=&\frac{e^{-2(a+b)\sigma }}{6\sigma^{2}}(bBe^{a\sigma}+aAe^{b\sigma})\nonumber\\
&\times&[Be^{a\sigma}(3+b\sigma)+e^{b\sigma}(A(3+a\sigma)+3e^{a\sigma}W_{0})].
\label{dspotential}
\end{eqnarray}
The resulting potential is shown in Fig.[\ref{KLp}]. With the values
of the parameters used in this figure, we may find a local minimum,
$V_{ds}$, at $\sigma =\sigma _{ds}\approx 62$ and an anti-de Sitter
global minimum, $V_{ads}$, at $\sigma =\sigma_{ads}\approx 106$. In
fact, the potential value at $\sigma_{ds}$ is negative. But if we
add the lifting term $\sim D/\sigma^{3}$ and fine tune this term, we
can always make the value of the local minimum, $V_{ds}$, be equal
to the observed cosmological constant $\Lambda\sim 10^{-120}$. (Of
course, the curve of the potential would be changed slightly. And
the global minimum at $\sigma_{ads}$ is still designed to be an AdS
vacuum.) From now on we just suppose this has been achieved.

\section{the Decay of the de Sitter Vacua}

Here we identify the dark energy in our universe as the de Sitter
vacuum of the volume modulus field, $V_{ds}$, and take other
components in the universe to be negligible. The metric of the
universe is the FRW metric
\begin{equation}
ds^{2}=-dt^{2}+a(t)^{2}(dr^{2}+r^{2}d\theta^{2}+r^{2}\sin
^{2}\theta d\varphi ^{2}).  \label{metric}
\end{equation}
The evolution of the universe is governed only by
\begin{equation}
H^{2}=\frac{1}{3M_{pl}^2}V_{ds},  \label{dsFriedmanEq}
\end{equation}
where we take $8\pi G=M_{pl}^{-2}$ and
$H=\frac{da}{dt}/a=\dot{a}/a$. From this equation, it is obvious
that the universe is at a de-Sitter state. But this would not last
for ever. As time goes by, due to the quantum mechanics, the decay
of the metastable dS vacuum state at $\sigma_{ds}$ is inevitable.
So, at some time, within some region, the volume modulus field
$\sigma$ tunnels through the barrier between the two minimum points
(as shown in Fig.[\ref{KLp}]), and appears at the AdS point,
$\sigma=\sigma_{ads}$. Notably, this means the modulus is still
stablized after the decay. It should be noted that this property is
absent in the simplest KKLT potential in \cite{KKLT}, where the
decay means the decompacting of the modulus $\rho$.

This vacuum decay may be treated as a first-order phase
transition. According to \cite{Coleman}, the tunnelling first
happens within a region and forms a bubble, and then the bound of
the bubble would expand outward at the speed of light. Outside of
the bubble, the universe is still at the dS phase. But inside, the
universe is at the AdS state. The metric becomes
\begin{equation}
ds^{2}=-dt^{2}+a(t)^{2}(\frac{dr^{2}}{1+r^{2}}+r^{2}d\theta ^{2}+r^{2}\sin
^{2}\theta d\varphi ^{2}).  \label{metricInside}
\end{equation}
The two metrics (\ref{metric}) and (\ref{metricInside}) are joined
at the bound of the bubble. If we choose the center of the bubble at
the moment of materialization to be the center of coordinates, the
bound of the bubble is at $r=\bar{r}(t)$. Both the initial value of
the bound, $\bar{r}(0)$, and the probability of the decay, can be
obtained by using the thin wall approximation \cite{Coleman}. But
here these values are of no interest to us. We just need verify that
the condition for the decay to happen \cite{Coleman},
\[
\frac{3S_{1}^{2}}{4(V_{ds}-V_{ads})}<1,
\]
is satisfied. Using the potential shown in Fig.[\ref{KLp}], we get
$\frac{3S_{1}^{2}}{4(V_{ds}-V_{ads})}\simeq 0.66$. So the condition
is satisfied.

\section{the Evolution after the Decay}

Now let us analyze the evolution of the universe after the decay. At
first, for cosmological purposes it is convenient to define the
canonical variable $\phi =\sqrt{\frac{3}{2}}\ln \sigma $. From now
on, we will use the field $\phi$, instead of $\sigma$.

Outside the bubble, everything is the same as before and the
evolution is still governed by the equation (\ref{dsFriedmanEq}).
But inside, the moving equation becomes
\begin{equation}
H^{2}=\frac{1}{a^{2}}-\Lambda_{a}^{2},  \label{adsFriedmanEq}
\end{equation}
where $\Lambda _{a}\equiv\sqrt{-\frac{1}{3M_{pl}^2}V_{ads}}$. This
equation has the solution
\begin{equation}
a(t)=\frac{\sin (\Lambda_{a}t+\delta )}{\Lambda_{a}},
\label{solutionInside}
\end{equation}
where $\delta$ is the integral constant. Taking account of the
expansion of the universe before the decay, we can conclude that
the universe in the bubble is expanding at the early stage after
the decay. This is ensured by the condition
\[
0<\delta <\pi /2.
\]
Obviously, from the solution (\ref{solutionInside}), the contraction
would happen after the scaling factor $a$ arrives its maximum
$a_{m}=\Lambda_{a}^{-1}$. Then the singularity of the big crunch
will be encountered at the end of the contraction.

Naively, if the singularity of the big crunch is neglected, it seems
that we can expect a cyclic model from Eq.(\ref{solutionInside}).
But, it is not the case. In fact, the solution
(\ref{solutionInside}) is not valid even during the whole process of
the first contraction. The reason is that this equation is an ideal
solution by neglecting the effect of the perturbation. In
\cite{Coleman}, we know, a very small velocity of $\phi$ towards the
false vacuum is the necessary condition for the instanton solution
to exist. Then, considering the perturbation, there must exist a
small value of the kinetic energy term of the field,
$\dot{\phi}^2/2$, in side the bubble. On the other hand we know
$\dot{\phi}^2/2\propto a^{-6}$. So, during the contraction, the
small kinetic energy increases rapidly and can not be neglected
after a while. Indeed, after the sufficiently large time, this
kinetic energy would dominates the part of the universe. Then the
evolving equation would be
\begin{equation}
\label{FridemanEqInOpen}H^{2}=\frac{1}{a^{2}}+\frac{1}{3}(\frac{1}{2}\dot{\phi}^{2}+V).
\end{equation}
The solution (\ref{solutionInside}) would be invalid.


However, according to the equation
\[
\dot{H}=-\frac{1}{2M_{pl}^2}(1+w)\rho -\frac{1}{a^{2}},\ p=w\rho
\]
we get that $a(t)$ continues to decrease even after the solution
(\ref{solutionInside}) invalid. Of course, in this paper we only
consider the component with $w\geq-1$. Then the singularity of the
big crunch seems to be inevitable.

\section{the Bounce of the Universe at the Planck Scale and the Cyclic Scenario}

We know that the energy scale of the universe increases as the
contraction. So the singularity of the big crunch is a problem about
the physics at the Planck energy scale. At the same time it is
generally believed that the big crunch singularity should not be a
feature of quantum gravity and there might exit some mechanism
to avoid the singularity. To solve this problem, many conjectures
have been suggested \cite{a0405353,g0312110,t0506218}. In this
section, to deal with it, we use the result shown in \cite{t0506218}
directly. In \cite{t0506218}, the Friedmann is modified as
\begin{equation}
H^{2}=\frac{1}{3M_{pl}^2}\rho(1-\frac{\rho}{M_{pl}^4}),
\label{modifiedFriedmannEq}
\end{equation}
where $\rho$ denotes the energy density in the universe. Of course
this equation is the modified Friedmann equation in the flat
universe. However, if the curvature term, $\frac{1}{a^2}$, in
(\ref{FridemanEqInOpen}) is also taken as a component in the
universe with the energy density $\rho=\frac{3M_{pl}^2}{a^2}$,
Eq.(\ref{modifiedFriedmannEq}) is applicable inside the
vacuum-decay bubble.

It has been shown in \cite{t0506218} that this modified Friedmann
equation avoids the catastrophe of the big crunch by giving a
bounce at $\rho=M_{pl}^2$. After the bounce, the universe begins
to expand. Naturally, we may take the bounce as the big bang.
Furthermore, we find that a cyclic cosmological model is obtained.
Now let's show it.

It has been given in the last section, as the bounce being
approached, the kinetic energy term of $\phi$,
$\frac{1}{2}\dot{\phi}^{2}$ scaling as $a^{-6}$, becomes the
dominant component of the universe. Even, the velocity of $\phi$ is
so large that the field can roll up and over the barrier of the
potential and continues rolling up along the potential. After the
bounce, the energy density of the modulus field, $\phi$, begins to
decrease as the expansion of the universe. The field, $\phi$ would
roll down along the potential, too. Eventually the field might stay
at the dS vacuum state, $V_{ds}$. Notably, the potential used here
is just a toy model. Actually, in the scenario of \emph{landscape},
it is argued that there exist many vacua. Then, after the bounce, it
is possible for the field to roll down to any vacuum. This means the
universe, after the bounce, may have a different cosmological
constant in each cycle \cite{t0407258}.

No losing the generality, we assume that, after the bounce, the
initial position of the field is appropriate and the rolling-down
field can not overshoot to pass barriers between vacua. Then there
two different cases after the bounce. One is that, after bounce, the
field rolls down to an anti-de Sitter minimum. In this case, the
contraction would happen during the oscillating of the field around
the minimum. The moving equation is Eq.(\ref{FridemanEqInOpen}).
During the expansion, the potential energy, $V(\phi)$, becomes
negative and the right-hand side would be zero as the decreasing of
the kinetic energy term, $\frac{1}{2}\dot{\phi}^2$, and the
curvature energy term, $\frac{1}{a^2}$ (Of course the curvature term
may be neglected). The turnaround from expansion to contraction is
obtained at $H=0$. Then the universe contracts and bounces to evolve
into the next new cycle.

The other case is that, after bounce, the field is rolling down to
an de Sitter minimum. In this case the evolution of the universe is
similar to our universe. The vacuum would decay into an anti-de
Sitter vacuum and then contract, bounce to evolve into another new
cycle. Notably, in this case, only one part of the universe in the
previous cycle, the part inside the vacuum-decay bubble, contracts,
bounces and then develops to be a new observed universe. Then the
initial entropy of the observed universe in the new cycle is only
one part of the entropy in the last cycle. If we suppose that it is
much more possible for the field to roll down to dS vacua than to
AdS vucua, the entropy of the observed universe would not become
larger and larger as the cycle repeating in our scenario. Of course,
the total entropy of the whole universe grows, in accord with the
second law of thermodynamics,

Now we can conclude that the contraction/expansion cycles are
always obtained in both cases, although there is a difference
between them. And the problem of the infinite entropy is avoided
in some sense.

\section{Summary and Discussion}

In this paper, we first assume that the dark energy is the
de-Sitter vacuum of the potential in \cite{KL} and then analyze
the decay of the metastable dS vacuum. We show that the decay of
the vacuum is inevitable and inside the vacuum-decay bubble, the
modulus field, $\sigma $, tunnels through the barrier of the
potential to appear at the anti-de Sitter vacuum state. We find
that the universe in the bubble would contract to the singularity.
By the modified Friedmann equation in \cite{t0506218}, the
singularity is avoided and a bounce appears as the end of the
contraction. Then we show that this give a cyclic cosmological
model.

This model is different from ordinary cyclic models
\cite{a0404480,t0403020}. In our scenario, the universe can
experience many cycles with different vacua. And the future
turnaround happens naturally due to the vacuum decay. This implies
another difference that only the part in the vacuum-decay bubble
contracts, then bounces and expand to be a new observable universe.
Of course, here our model is only a toy model. But we believe it is
significant to incorporate the \emph{Landscape} scenario with the
cyclic cosmological model. However, the details of our mechanism
need to be explored further. There is much work to do in order to be
sure this cyclic scenario to work well.

\end{document}